\documentstyle[12pt]{article}
\def\AP#1#2#3{{\it Ann.\ Phys.} {\bf #1}, #2 (#3)}

\def\PRD#1#2#3{{\it Phys.\ Rev.} {\bf D#1}, #2 (#3)}
\def\PRL#1#2#3{{\it Phys.\ Rev.\ Lett.} \ {\bf #1}, #2 (#3)}
\def\NPB#1#2#3{{\it Nucl.\ Phys.} {\bf B#1}, #2 (#3)}

\def\PLB#1#2#3{{\it Phys.\ Lett.} {\bf B#1}, #2 (#3)}

\def\PRP#1#2#3{{\it Phys.\ Rep.} {\bf #1}, #2 (#3)}

\def\IJMPA#1#2#3{{\it Int.\ J.\ Mod.\ Phys.} {\bf A#1}, #2 (#3)}

\def\MPLA#1#2#3{{\it Mod.\ Phys.\ Lett.} {\bf A#1}, #2 (#3)}

\def\ANP#1#2#3{{\it Ann.\ Physics (N.Y.)} {\bf #1}, #2 (#3)}
\def\TMP#1#2#3{{\it Theor.\ Math.\ Phys.} {\bf #1}, #2 (#3)}
\newcommand{\be}{\begin{equation}}
\newcommand{\ee}{\end{equation}}
\newcommand{\ba}{\begin{eqnarray}}
\newcommand{\ea}{\end{eqnarray}}
\def\d{\partial}
\begin{document}
\title{Two-Dimensional Reduced Theory and General Static Solution for 
Uncharged Black $p$-Branes}
\author{Marco Cavagli\`a\thanks{New address:
Max-Planck-Institut f\"ur Gravitationsphysik,
Albert-Einstein-Institut,
Schlaatzweg 1,
14473 Potdsam, Germany}
\\ \it
Department of Physics and Astronomy \\ \it
Tufts University \\ \it
Medford, MA 02155 U.S.A.}
\maketitle

\begin{abstract}

We derive a two-dimensional effective dilaton - gravity - matter action
that describes the dynamics of an uncharged black $p$-brane in $N$
dimensions. We show that this effective theory is completely integrable in
the static sector and establish its general static solution. The solution
includes, as a particular case, the boost symmetric $p$-brane solution
investigated in Ref.\ \cite{ruth}.

\end{abstract}

\section{Introduction}

Recently a large amount of time has been devoted to the investigation of
solutions of Einstein and/or dilaton gravity in dimensions other than
four \cite{nota1}.

In lower dimensional gravity the study of solutions in two- and three-
dimensions has received a lot of attention because of its connection
with string theory \cite{witt}, dimensionally reduced models
(minisuperspaces and midisuperspaces) \cite{minimidi} and the black hole
physics \cite{bh2d}. Lower dimensional models may further provide some
insight into the difficult challenge of quantizing gravity theories in the
(more physical) four dimensional case \cite{nico}. Hence, many 0+1, 1+1
and 2+1 integrable models have been analyzed and solved in the literature,
both from the classical and quantum points of view \cite{nico, q2d}. 

On the other hand, the investigation of solutions in more than four
dimensions has also raised considerable interest \cite{mel, mel2}. The
most promising candidate for a unified theory of all interactions, the so
called M-theory, is indeed naturally formulated in a $N>4$ dimensional
spacetime so it seems natural to query how extra dimensions are affecting
the four-dimensional world; it is well know, for instance, that in $N>4$
dimensions event horizons may have interesting non trivial topologies
\cite{ruth}. 

In this context many papers have been devoted to the discussion of
uncharged and charged (black) $p$-branes \cite{ruth, stelle, duff}, that
are essentially a D-dimensional black hole times a $p$-dimensional flat
Euclidean space. The analysis of black $p$-branes is indeed 
relevant both to supergravity theories \cite{stelle} and to cosmology, 
since $p$-branes can also be seen as topological defects \cite{ruth}. 

The aim of this paper is to give a contribution to the discussion of both
lower dimensional models and pure gravity $p$-brane solutions. Starting
from a generic ansatz in $N$ dimensions with topology D-hole $\times$
$p$-brane we reduce the $N$ dimensional Einstein-Hilbert action to a
two-dimensional dilaton gravity action plus a massless scalar field with a
non-trivial coupling to the dilaton. The static sector of this effective
theory turns out to be completely integrable and we establish the general
static solution of the system. We believe that the discussion of this
model is important at least for two reasons: first it represents an
example of an integrable dilaton - gravity - matter model in 1+1
dimensions with a non trivial coupling between the dilaton and the scalar
field. To our knowledge there are no other integrable models with these
characteristics in the literature \cite{fil}. Further, the general static
solution of this system is also the general static solution in $N$
dimensions of a generic uncharged black $p$-brane with dimensions $p\le
N-2$.  This solution includes, as a particular case, the generic uncharged
solution with boost symmetry of Ref.\ \cite{ruth}. We will also see that
there are indeed two sets of boost symmetric $p$-branes corresponding to the
solutions of Ref.\ \cite{ruth} and that these are related by a
duality-like relation. 

The paper is thus divided in two steps. We first derive the effective
two-dimensional action and write the equations of motion. Then we
integrate the equations of motion in the static sector and discuss their
connection with the boost symmetric solutions of Ref.\ \cite{ruth}. 

\section{Two-Dimensional Effective Theory}

Our starting point is the Einstein-Hilbert Lagrangian in $N\ge 4$
dimensions. We consider a multidimensional manifold endowed with a
pseudo-Riemannian metric of the form
\be 
ds^2=g_{\mu\nu}(y)dy^\mu dy^\nu+f(y)\delta_{ij}dz^idz^j\,,
\label{m1}
\ee
where $\mu,\nu=0,...N-p-1$, $i,j=N-p,...N-1$, and $g_{\mu\nu}$ is a
pseudo-Riemannian spherically symmetric metric in $N-p$ dimensions. The
ansatz (\ref{m1}) includes as a particular case the boost
symmetric metric of a $p$-brane of Ref.\ \cite{ruth}

The model can be reduced to a two-dimensional effective theory using a 
suitable ansatz for the line element (\ref{m1}). Write (\ref{m1}) in the form
\be
ds^2=\phi^{{1\over p+q}-1}h_{ab}(x)dx^a 
dx^b+\phi^{{2\over p+q}}e^{\psi}d\Omega^{(q)}
+\phi^{{2\over p+q}}e^{-{q\over p}\psi}\delta_{ij}dz^idz^j\,, 
\label{m2}
\ee
where $\phi$ and $\psi$ are two functions of $x$, $d\Omega^{(q\equiv
N-p-2)}$ is the metric of the $q$-dimensional unit sphere, and $h_{ab}$
($a=0,1$) is a generic two-dimensional pseudo-Riemannian metric. 
Using (\ref{m2}) the $N$-dimensional Einstein-Hilbert action reduces to the 
form (we neglect surface terms that are irrelevant for the present 
discussion)
\be
S={1\over\kappa}\int d^2x\sqrt{-h}\left[\phi 
R^{(2)}(h)+Y(\phi,\psi)+Z(\phi)(\nabla\psi)^2\right]\,,
\label{action}
\ee
where $R^{(2)}(h)$ is the two-dimensional Ricci scalar and $Y$ and $Z$ 
are the two functionals
\be
Y(\phi,\psi)=q(q-1)e^{-\psi}\phi^{-{1\over 
p+q}}\,,~~~~~~~~Z(\phi)=-{q\over 4p}(p+q)\phi\,.
\label{terms}
\ee
$\kappa$ is a constant factor equal to $16\pi G/(V_b^{(p)}V_{s}^{(q)})$ 
where $V_b^{(p)}$ and $V_s^{(q)}$ are respectively the volume of the
unit $p$-brane and of the unit $q$-sphere. Without loss of generality, we 
will set $\kappa=1$ and thus in the following (\ref{action}) must be 
considered an action density in $p$-brane space.

The theory described by the action (\ref{action}) is an example of a
two-dimensional dilaton - gravity - matter model with a non trivial
coupling between the matter field and the dilaton field since $\d_\phi
Z\not=0$ \cite{fil}. All fields in (\ref{action}) represent gravitational
degrees of freedom of the $N$-dimensional spacetime. The model can be of
course extended to include a genuine matter field, as the dilaton, and the
Maxwell field. For instance, working in the Einstein frame the dilaton
term is
\be
S_d=\int d^2x\sqrt{-h}\left[-{4\over p+q}\phi(\nabla\chi)^2\right]\,,
\label{dil}
\ee
where, of course, the dilaton field $\chi$ depends only on the $x$
coordinates. The dilaton contributes to the action essentially as the
field $\psi$. Solving the equation of motion for the dilaton, one can see
that its contribution reduces to a potential term $V(\phi)$ in the action
(\ref{action}). Thus the inclusion of the dilaton, as well as the Maxwell
field, modifies the dynamics of the system with respect to the pure
gravity case. The discussion of these cases is however beyond the limited
purposes of this report, so we refer the reader to a future paper on the
subject and concentrate now on the pure gravity case. 

>From the action (\ref{action}) it is straightforward to obtain the equations
of motion. They can be cast in a useful and simple form writing the 
two-dimensional metric $h_{ab}$ in the ``conformal gauge'' \cite{fil}
\be
ds^2_{h}\equiv h_{ab}dx^adx^b=4\rho(u,v)dudv\,.
\label{mconf}
\ee
The result is
\ba
&&\d_u\d_v\phi-\alpha\rho\, e^{-\psi}\phi^{\beta}=0\,,
\label{ephi}\\ \nonumber\\
&&\d_u\d_v(\ln\rho)-\alpha\beta\rho\, 
e^{-\psi}\phi^{\beta-1}-\gamma\d_u\psi\d_v\psi=0\,,
\label{erho}\\ \nonumber\\
&&\gamma\d_u(\phi\d_v\psi)+\gamma\d_v(\phi\d_u\psi)+\alpha\rho\, 
e^{-\psi}\phi^\beta=0\,,
\label{epsi}\\ \nonumber\\
&&\rho\d_u\left({\d_u\phi\over\rho}\right)=\gamma\phi(\d_u\psi)^2\,,~~~~
\rho\d_v\left({\d_v\phi\over\rho}\right)=\gamma\phi(\d_v\psi)^2\,,
\label{constr}
\ea
where $\alpha=q(q-1)$, $\beta=-1/(p+q)$, and $\gamma=-q(p+q)/4p$. The
equations of motion (\ref{ephi}-\ref{epsi}) and the constraints
(\ref{constr}) form a set of five non-linear and coupled second order
partial differential equations in the unknown $\rho$, $\phi$, and $\psi$. 
However, as remarked in Ref.\ \cite{fil} only four of these equations are
independent because Eqs.\ (\ref{ephi},\ref{erho}) and the constraints
(\ref{constr}) imply (\ref{epsi}), or, alternatively, Eqs.\ 
(\ref{ephi},\ref{epsi}) imply (\ref{erho}). 

Due to the nature of the system, the general solution of the equations of
motion is very hard to find. However, it is possible to find the general
static solution associated to the system of equations
(\ref{ephi}-\ref{constr}). According to Filippov \cite{fil} we define
``static'' those solutions of the equations of motion that can be cast in the
form
\be
\psi\equiv\psi(\tau)\,,~~~~~
\phi\equiv\phi(\psi)\,,~~~~~
\rho\equiv\xi(\psi)\d_u\tau\d_v\tau\,,
\label{static}
\ee
where $\tau$ is a harmonic function: $\d_u\d_v\tau(u,v)=0$. The solutions
(\ref{static}) depend only on one coordinate ($\tau$). This also implies
that the metric coefficients in (\ref{m2}) depends on a single coordinate,
and thus the geometry is static. In particular, since $\tau$ is harmonic
we can write $\tau=U(u)+V(v)$, and from (\ref{mconf}) we see that $U\mp V$
can be identified with the timelike coordinate and with the conformal
radial coordinate in the line element (\ref{m2}) respectively. The
arbitrariness in the choice of $U(u)$ and $V(v)$ reflects of course the
invariance of (\ref{mconf}) under coordinate reparametrization.

In the calculation of static solutions we are guided by the case of
two-dimensional pure dilaton - gravity models that has been proved to be
completely integrable for any dilaton potential \cite{fil}. In that case
it is also proved that all solutions are of the static type (extended
Birkhoff theorem). In the present case we are only able to prove that the
static sector is completely integrable and find the general static
solution. It would  then be worthwile to investigate if these solutions are 
also the general (static and non-static) solutions of the system, extending the
Birkhoff theorem to the (black) $p$-branes (\ref{m2}).

Substituting (\ref{static}) in the Eqs.\ (\ref{ephi}-\ref{constr}), these
reduce to the form
\ba
&&{d^2\phi\over d\psi^2}\psi'^2+{d\phi\over 
d\psi}\psi''-\alpha\xi e^{-\psi}\phi^{\beta}=0\,,
\label{ephi2}\\ \nonumber\\
&&
{d^2~\over d\psi^2}(\ln\xi)\psi'^2+{d~\over
d\psi}(\ln\xi)\psi''-\alpha\beta\xi
e^{-\psi}\phi^{\beta-1}=\gamma\psi'^2\,,
\label{erho2}\\ \nonumber\\
&&2\gamma{d\phi\over d\psi}\psi'^2+2\gamma\phi\psi''+\alpha\xi
e^{-\psi}\phi^\beta=0\,,
\label{epsi2}\\ \nonumber\\
&&{d^2\phi\over d\psi^2}\psi'^2+{d\phi\over
d\psi}\psi''-{d\phi\over
d\psi}{d~\over d\psi}(\ln\xi)\psi'^2=\gamma\phi\psi'^2\,,
\label{constr2}
\ea
where primes represent derivatives with respect to $\tau$.  Using the
equations of motion in the form above it is straighforward to obtain two
first integrals and further reduce the system. After some algebra one
obtains
\ba
&&\phi{d^2\phi\over d\tau^2}+a\left({d\phi\over 
d\tau}\right)^2+b{d\phi\over d\tau}=c\,, \label{phi} \\ \nonumber \\
&&\phi'+2\gamma\phi\psi'=K\,, \label{kappa} \\ \nonumber \\
&&\xi=\xi_0\phi^{1+\beta+\delta}e^{2\gamma\delta\psi}\,,\label{xi} \\
\nonumber \\
&&{d^2\phi\over 
d\tau^2}=\alpha\xi_0\phi^{1+2\beta+\delta}e^{(2\gamma\delta-1)\psi}\,, 
\label{phipsi}
\ea
where $a=-(1+\beta+1/4\gamma)\equiv (1-q)/q$, $b=-K(\delta-1/2\gamma)$, and
$c=K^2/4\gamma$. $\xi_0$, $K$, and $\delta$ are three constants of motion.
The first one has no physical meaning and can be reabsorbed in the
definition of $\tau$. The other two play a fundamental role in the
geometry of the $p$-brane as we will see below. This set of equations can
be read as follows: the integration of Eq.\ (\ref{phi}) gives $\phi$, the
solution of Eq.\ (\ref{kappa}) defines $\psi$ once $\phi$ is known,
finally Eq.\ (\ref{xi}) defines the conformal part of the two-dimensional
metric $\xi$ as a function of $\phi$ and $\psi$; Eq.\ (\ref{phipsi}) is
dependent from the others and is used to determine the constants of 
integration. 

\section{Static Black $p$-Branes}

The system of Eqs.\ (\ref{phi}-\ref{phipsi}), and in particular Eq.\
(\ref{phi}), is the starting point to solve the system. Before looking for
the general solution, let us calculate as a warming up exercise the
boost symmetric solutions of Ref.\ \cite{ruth}. 

\subsection{Boost Symmetric $p$-Branes}

In order to compare our results with those of Ref.\ \cite{ruth} let us
look for a solution for $\phi$ of the the form $\phi=r^m A(r)^n$, where
$m$ and $n$ are real numbers to be determined by the equations of motion,
and $r$ is related to $\tau$ by the relation $\tau=\int dr A(r)^{n-1}$.
Choosing $\delta=-2\beta$ (boost symmetric $p$-brane, see Eq.\ (\ref{m2}))
and after some algebra Eq.\ (\ref{phi}) can be solved and the system
integrated.  The final result is
\ba
&&\phi=r^q A^{{q(\Delta-1)\over 2\Delta(q-1)}}\,,\\
&&e^{\psi}=r^{{2p\over
p+q}} A^{
{p(\Delta-q)\over\Delta(q-1)(p+q)}}\,,\\
&&\xi=\phi^{1+{1\over p+q}}e^{-{q\over p}\psi}\,,~~~~~q>1\,,
\ea
where 
\be
A(r)=1-\left({r_+\over r}\right)^{q-1}\,,
\label{a}
\ee
and $\Delta$ and $r_+$ are two constants defined by the relations
\be
\Delta^2=q(p+1)/(p+q)\,,~~~~~K=-{q(q-1)\over 2\Delta}r_+^{q-1}\,.
\label{Delta}
\ee
Finally, the boost symmetric $p$-brane reads 
\be
ds^2=A^{{\Delta\over p+1}}(-dt^2+dz_i^2)+A^{{2-q-\Delta\over 
q-1}}dr^2+r^2A^{{1-\Delta\over q-1}}d\Omega^{(q)}\,,
\label{boosted}
\ee
Several interesting remarks can be suggested by the solution
(\ref{boosted}).  First of all it is immediate to see that $K$ introduced
in Eq.\ (\ref{kappa}) is related to the radius of the horizon of the black
$p$-brane (\ref{boosted}). The role of the constant of motion $K$ is thus
analog to the role of the mass in two-dimensional dilaton - gravity
models, and in particular in the effective two-dimensional dilaton - 
gravity theory describing the Schwarzschild black hole.
In those cases the existence of a functional conserved under time
and space translations (the mass) is strictly related to the validity of
the generalized Birkhoff theorem and to the dimensional collapse of the
$1+1$ theory into a $0+1$ theory (see e.g.\ \cite{birkh}). In the 
present case
the existence of $K$ seems to suggest a similar conclusion, i.e.\ that all
the solutions of the model described by (\ref{action}) are actually
static. Second, in (\ref{boosted}) there are two different sets of
solutions related by the sign of $\Delta$. The (uncharged) boost symmetric
$p$-branes of Ref.\ \cite{ruth} correspond to the choice of positive
$\Delta$. However, if one performs the discrete ``dual'' transformation
$\Delta\to -\Delta$ other solutions are obtained.  For instance in the
case of the 5-brane in 10 dimensions we obtain the two solutions
\be
ds^2=\left(1-{r_+^2\over r^2}\right)^{n_1}
(-dt^2+dz_i^2)+\left(1-{r_+^2\over r^2}\right)^{n_2}
dr^2+r^2\left(1-{r_+^2\over r^2}\right)^{n_3}d\Omega^{(3)}\,,
\label{boost10}
\ee
where $n\equiv(1/4,-5/4,-1/4)$ for $\Delta=3/2$ and $n\equiv(-1/4,1/4,5/4)$ 
for $\Delta=-3/2$. The ``duality'' invariance $\Delta\to -\Delta$ changes 
in particular the radius of the boost symmetric $p$-brane into its inverse. 
From
the definition of $\Delta$ (\ref{Delta}) it is straightforward to see 
that $\Delta^2>1$ for any value of $p$ and $q$ ($>1$, of course). This 
implies that the singularity of the metric at $r=r_+$ has a divergent 
volume element for the set of solutions $\Delta>0$ and a finite volume 
element for the set with negative $\Delta$.
 
Finally, let us stress that the parameter $\delta$ defined in Eqs.\
(\ref{ephi2}-\ref{constr2}) has been chosen equal to $-2\beta$. Boost 
symmetric $p$-branes represent thus only a set of null measure in the space 
of static solutions.

\subsection{General Static Solution}

Let us now move to the calculation of the general static solution. We 
will see that the structure of the space of general static solutions is 
much richer that the space of the boost symmetric solutions.

The system (\ref{phi}-\ref{phipsi}) can be exacly integrated without
introducing the semplifications used in the previous calculation.  This
program can be completed defining a new coordinate $x$ related to $\tau$
by the expression $\tau=\int \phi(x)dx$. After some algebra the general
solution is found to be
\ba
&&\phi=C_\phi e^{-{K\over 4a\gamma}(1-2\gamma\delta)x}
\left(\sinh{(K\bar\Delta x)}\right)^{1\over a}\,,\\
&&e^\psi=C_\psi e^{-{K\over 4a\gamma}[\delta+2(1+\beta)]x}
\left(\sinh{(K\bar\Delta x)}\right)^{-{1\over 2a\gamma}}\,,\\ 
&&\xi=C_\xi  e^{-{K\over 
4a\gamma}[\delta+(1+\beta)(1+2\gamma\delta)]x}
\left(\sinh{(K\bar\Delta x)}\right)^{{1+\beta\over a}}\,,
\ea
where $C_\xi 
C_\psi^{-1}C_\phi^{1+\beta}=x_0^{-2}\equiv K^2\bar\Delta^2/(q-1)^2$ and
\be
\bar\Delta^2={1\over 
4}\left[\delta\left(\delta-{1\over\gamma}\right)-
{1+\beta\over\gamma}\right]\,.
\label{delbar}
\ee
The metric of the $p-$brane reads then
\ba
ds^2=-A_1^2e^{K\delta x}dt^2+A_2^2 e^{-2\beta Kx}dz_i^2+
e^{-{K\over 2a\gamma}[1+2\gamma\delta(\beta+1/4\gamma)]x}\cdot\nonumber\\
\cdot A_3^2\sinh^{{2\over a}}{(K\bar\Delta x)}
\left[dx^2+x_0^2\sinh^2{(K\bar\Delta x)}d\Omega^{(q)}\right]\,,
\label{genbrane}
\ea
where $A_1^2=C_\xi C_\phi^{-(1+\beta)}$, $A_2^2= (C_\xi
C_\phi^{1+\beta+1/2\gamma}x_0^2)^{-4\beta\gamma}$, $A_3^2=C_\xi
C_\phi^{1-\beta}$, and the first two may be eventually reabsorbed in the
redefinition of the time and $p$-brane coordinates. It is straightforward
to check that, setting $\delta=-2\beta$, (\ref{genbrane}) becomes the boost
symmetric solution of Sec.\ 3.1. This can be easily proved using the 
coordinate transformation
\be
x={q\over 2K\Delta}\ln\left[1-\left({r_+\over r}\right)^{q-1}\right]\,,
\label{ctrasf}
\ee
and setting $A_3=2^{q/(1-q)}r_+^q$. In this case $x_0^2=4r_+^{2(1-q)}$,
$\bar\Delta^2=\Delta^2/q^2$, and the ``duality'' relation of the previous
section is immediate since the metric is evidently invariant under change
of sign of $\Delta$. 

Starting form the line element (\ref{genbrane}), the geometrical
properties of the general solution can be investigated for any value of
the parameter $\delta$. This is beyond the scope of this short note, so we
will not enter into details. Let us stress, however, that all solutions
(\ref{genbrane}) are asymptotically flat for $x\to 0$, corresponding to
$r\to\infty$ in the boost symmetric case of Sec.\ 3.1, for any value of
the parameter $\delta$. 

\section{Conclusions}

The effective two-dimensional dilaton - gravity - matter action
(\ref{action}) describes the dynamics of uncharged $p$-branes in $N\ge 4$
dimensions. In this paper we have proved that the model is completely
integrable in the static sector and we have calculated the general
solution. This result can be read from two different perpectives. First it
is an example of an integrable model of two-dimensional dilaton gravity
plus matter (at least in the static sector) with a non trivial coupling
between the fields, and the knowledge of exact solutions of complex
theories is always welcome in theoretical physics. Second, the general
solution of Sec.\ 3.2 describes the most general static uncharged black
$p$-brane in $N\ge 4$ dimensions. The existence of the general static
solution of Sec.\ 3.2, and its relation to the two-dimensional theory
(\ref{action}), imply that some essential properties of (uncharged)
$p$-branes are very similar to the properties of solutions of other
well-known dilaton - gravity models, as for instance the spherically
simmetric reduced pure gravity. Among these properties, the definition of
the $p$-brane mass in canonical form and the extension of the Birkhoff
theorem to this class of theories (see e.g.\ \cite{birkh}), are certainly
worth being explored in the future. 

\section{Acknowledgements}

This work was in part supported by the foreign grant N.\ 3229/96 of the
University of Torino and by the Angelo Della Riccia Foundation, Florence,
Italy. We are very grateful to Ruth Gregory, Paulo Moniz, Alex Vilenkin,
and, in particular, to Vittorio de Alfaro for interesting discussions and
useful suggestions on various questions connected to the subject of this
paper.


\thebibliography{999}

\bibitem{nota1}{It is very difficult to give a detailed and complete
bibliography covering all aspects of this subject. The references cited
below are intended to be nor a complete nor an exhaustive review of the
present research in classical and quantum gravity in generic $N$
dimensions. For a full discussion on the many aspects of this subject the
reader is referred to the excellent recent reviews cited in the
introduction. Other references about the most advanced research can also
be found in the recent paper by V.D.\ Ivashchuk, V.N.\ Melnikov, and M.\
Rainer \cite{mel} and in several contributions to the {\it Proceedings of the
 2nd Conference on Constrained Dynamics and Quantum Gravity}, Santa
Margherita Ligure, September 1996 (Proc. Supplement, Nucl. Phys. B, in press,
1997, eds.\ V.\ de Alfaro et al.. We apologize in advance for any omission.}

\bibitem{mel}{V.D.\ Ivashchuk, V.N.\ Melnikov, and M.\ Rainer,
``Multidimensional $\sigma$-models with composite electric $p$-branes'',
e-Print Archive:  gr-qc/9705005 and references therein.}

\bibitem{witt}{E.\ Witten, \PRD{44}{314}{1991}.}

\bibitem{minimidi}{H.\ Kastrup and T.\ Thiemann, \NPB{425}{665}{1994}; 
D.\ Marolf, \PRD{53}{6979}{1996}; K.V.\ Kucha\v{r}, \PRD{50}{3961}{1994}.}

\bibitem{bh2d}{C.\ Callan, S.\ Giddings, J.\ Harvey, and A.\ Strominger,
\PRD {45}{1005}{1992}; J.\ Russo, L.\ Susskind, and L.\ Thorlacius,
\PRD{46}{3444}{1992}; S.\ Carlip, J.\ Gegenberg, and R.B.\ Mann,
\PRD{51}{6854}{1995};  A.\ Barvinskii and G.\ Kunstatter,
\PLB{389}{231}{1996} and references therein.}

\bibitem{nico}{See for instance: H.\ Nicolai, D.\ Korotkin, and H.\
Samtleben, ``Integrable Classical and Quantum Gravity'', Lectures given at
{\it NATO Advanced Study Institute on Quantum Fields and Quantum Space
Time}, Carg\`ese, France, 22 July - 3 August.}

\bibitem{q2d}{E.\ Benedict, R.\ Jackiw, and H.-J.\ Lee,
\PRD{54}{6213}{1996}; D.\ Cangemi, R.\ Jackiw, and B.\ Zwiebach,
\ANP{245}{408}{1995}; D.\ Cangemi and R.\ Jackiw, \PRL {69}{233}{1992};
K.V.\ Kucha\v{r}, J.D.\ Romano, and M.\ Varadarajan, \PRD{55}{795}{1997};
D.\ Louis-Martinez, J..\ Gegenberg, and G.\ Kunstatter, \PLB
{321}{193}{1994} and references therein.}

\bibitem{mel2}{K.A.\ Bronnikov and V.N.\ Melnikov, \AP{239}{40}{1995};  V.D.\
Ivashchuk and V.N.\ Melnikov, \TMP{98}{212}{1994}.}

\bibitem{ruth}{R.\ Gregory, \NPB{467}{159}{1996}.}

\bibitem{stelle}{K.S.\ Stelle, ``Lectures on Supergravity $p$-branes'',
{\it Lectures presented at 1996 ICTP Summer School}, Trieste, Italy.}

\bibitem{duff}{M.J.\ Duff, H.\ Lu, C.N.\ Pope, \PLB{382}{73}{1996}; M.J.\
Duff, Ramzi R.\ Khuri, J.X.\ Lu, \PRP{259}{213}{1995}.}

\bibitem{fil}{A.T.\ Filippov, \MPLA {11}{1691}{1996}; \IJMPA{12}{13}{1997}.}

\bibitem{birkh}{M.\ Cavagli\`a, V.\ de Alfaro and A.T.\ Filippov, ``The
Birkhoff Theorem in the Quantum Theory of Two-Dimensional Dilaton
Gravity'', Report No.\ DFTT 20/97, e-Print Archive: hep-th/9704164.}
\end{document}